\documentclass[AMA,Times1COL]{WileyNJDv5} 

\articletype{Review}%

\received{Date Month Year}
\revised{Date Month Year}
\accepted{Date Month Year}
\journal{Journal}
\volume{00}
\copyyear{2023}
\startpage{1}

\usepackage{multirow}
\usepackage{graphicx}
\usepackage{array}
\usepackage{rotating}
\usepackage{vcell}
\usepackage{tabularray}
\usepackage{makecell}

\newcommand{\Y}{\mathbf{Y}}
\newcommand{\X}{\mathbf{X}}
\newcommand{\y}{\mathbf{y}}

\begin{document} 
\author[1]{Oliver Church}
\author[2]{Christopher Jackson}
\title{Modelling multiple disease risk factors for microsimulation studies: a review of methods}
\authormark{Church and Jackson}
\titlemark{Modelling multiple disease risk factors for microsimulation studies: a review of methods}
\address[1]{Work carried out while at MRC Biostatistics Unit, University of Cambridge}
\address[2]{MRC Biostatistics Unit, University of Cambridge}
\corres{Corresponding author: Christopher Jackson \email{chris.jackson@mrc-bsu.cam.ac.uk}}

\abstract[Abstract]{Longitudinal microsimulation models to evaluate policies and scenarios for chronic disease risk reduction typically involve simulating multiple risk factors for a synthetic population over time.  Various statistical methods have been used to accomplish this, but the principles behind them have never been comprehensively reviewed or critically compared.  We describe and review the range of methods that have been used to simulate multiple risk factors over time, given either longitudinal or repeated cross-sectional data, highlight those which are most practically applicable and flexible, and introduce some useful variants of them.  We describe procedures that can be used to check and compare these methods in practice based on fit to data, and present a case study where the methods are used and compared, based on a microsimulation model to evaluate mid-life cardiovascular health checks.  We conclude that a useful range of procedures are available, and this paper improves the knowledge base to support their use in models to inform health policy.}

\keywords{microsimulation,longitudinal,multivariate,cardiovascular}

\maketitle

\section{Introduction}

Longitudinal microsimulation is a technique that is increasingly used to evaluate policies to reduce the risk of chronic diseases \citep{briggs}. This involves simulating realistic trajectories of multiple risk factors and disease outcomes, at an individual level, for a synthetic population of interest.  The simulation is repeated under different policies or scenarios where particular risk factors are modified.  This allows the effect of policies or scenarios on population outcomes to be determined.  Outcomes are generally long-term, and may include estimated disease risk, disease incidence, mortality or health service costs.  Microsimulations are used for complex policies and long-term outcomes for which empirical studies such as randomised trials would be infeasible, and they usually combine evidence from many sources of data.  Simulating at the individual level can also capture variations in outcomes between subsets of the population \citep{brennan}.

\begin{figure}
  \centering
  \includegraphics[width=0.6\linewidth]{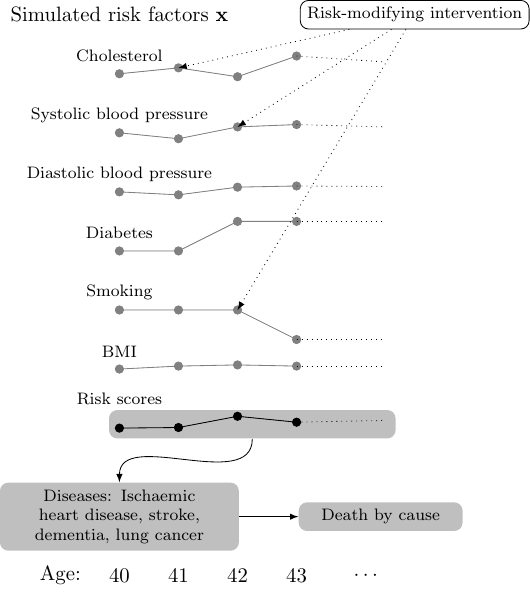}
  \caption{Illustration of the microsimulation model from Mytton et
    al~\citep{mytton} that assessed the effect of a health checks
    programme in England.  Risk factors for cardiovascular and other
    diseases are simulated year by year for a set of synthetic
    individuals, starting at age 40.  At each year, a risk score
    computed from the risk factors (e.g. QRisk3 for cardiovascular
    disease), which determines whether they get a disease, which they
    may later die from.  If a person attends a health check, they may
    receive treatments which may modify one or more risk factors.}
  \label{fig:microsim}
\end{figure}

An example of a longitudinal microsimulation model, used as a case study in this paper, is one that evaluated changes in a national mid-life health check programme \citep{mytton}.  In this programme, people over 40 years old are invited intermittently for a health check, where risk factors are measured.  Their overall 10 year risk of heart attack and stroke is described by the QRisk3 score~\citep{qrisk,qrisk3}. People at high risk might be recommended treatments, such as statins to lower cholesterol, or help with stopping smoking.   In the model, briefly, baseline data on risk factor progression from before the health checks programme were combined with observational data on health check delivery and rates of treatment following health checks, and data from randomised trials on the effects of treatments on risk factors.  Incidence of diseases and mortality are simulated given risk factors, via the risk scores.  The general model structure is illustrated in Figure~\ref{fig:microsim}.

A core part of a model like this is to simulate multiple disease risk factors for individuals over time (often on the scale of age, this typically being a main risk factor for chronic diseases).  The simulation is informed by longitudinal or cross-sectional data covering a range of ages.   Various different statistical methods have been used for simulating risk factors in this context.   In the health checks model, for example, key modelled risk factors include cholesterol (LDL and total), blood pressure (systolic and diastolic), smoking status and body mass index --- a mixture of continuous and discrete variables.   The simulation was done by an ad-hoc nonparametric procedure of resampling risk factor values from observed datasets: a ``baseline'' population of 40-44 year olds was obtained from census and survey data, and progressed forward by drawing values from matched longitudinal data.  Kingston et al. \cite{kingston} and May et al. \cite{may_cvd} simulated data from a Markov model with states defined by discretised risk factors, and transition probabilities estimated from longitudinal data.  Breeze et al. \cite{breeze2016} used random effects models for longitudinal data.  Lim et al. \cite{lim}, Kypridemos et al. \cite{kypridemos} and others used cross-sectional health survey data by age, and obtained longitudinal data via an assumption that individuals' risk factor ranks, relative to other individuals, were stable over time. Suen et al. \cite{suen} developed a variant of this procedure intended to better represent correlation between different variables.   

The statistical principles behind these various approaches, their relative theoretical advantages, and ways to empirically check and compare them in practice, have not been previously described in the context of longitudinal microsimulation.  This paper gives a review of methods that have been used for generating synthetic longitudinal risk factor data in health microsimulation, which use longitudinal or cross-sectional data or both.  We compare their principles and potential advantages, and describe how researchers can use to compare these methods given their own data.   We include variants of these methods that have not been used before in this context, e.g. the use of decision trees as a flexible approach to regression, and parametric models that allow the variance (as well as the mean) of risk factors to depend on predictors.  The ideas are illustrated using the data and risk factors from the health checks model --- which might plausibly be used in microsimulations for other chronic disease prevention questions.  The predictions of risk factors from the different simulation procedures are assessed by comparing against observed data.

After introducing the risk factor simulation model framework in Section~\ref{sec:overview}, we describe methods for simulating cross-sectional data at a single time in Section~\ref{sec:cross}.  The multivariate modelling techniques introduced there form the basis of methods to simulate longitudinal data given observed longitudinal data (Section~\ref{sec:long}) and methods to simulate longitudinal data given repeated cross-sectional data by age (Section~\ref{sec:long_csd}).  We discuss methods for checking and comparison of models for risk factors in Section~\ref{sec:checking}.  Sections \ref{sec:outcomes} and \ref{sec:interventions} briefly review how the simulation of disease outcomes and interventions, respectively, relates to the simulation of risk factors in microsimulations.  We illustrate a range of these ideas on the health checks case study in Section~\ref{sec:casestudy}.  We conclude with a discussion covering, e.g. extensions to longer term outcomes and combinations of data.

\section{Microsimulation of longitudinal risk factors: overall framework }
\label{sec:overview}

The aim is to simulate risk factor data for $R$ risk factors of interest, at a sequence of times or ages, $t \in \big\{ t_0=0,t_1,t_2,\ldots\big\}$ for a set of synthetic individuals $i$.  Denote the simulated data by $\y_i(t) =  \{y_{i1}(t),y_{i2}(t),...,y_{iR}(t) \}$ for each $t=t_j$, and the underlying time-varying random variable by $\Y(t)$. The simulation is done in the following steps.

\begin{itemize}
\item[(a)] Generate values of $\Y_i(0)$, and any time-constant characteristics $\X_i$ of individual $i$, obtaining the ``baseline'' population of interest at time $t=0$.  This is done using observed cross-sectional data (Section~\ref{sec:cross}).

\item[(b)] For each individual $i$, and each time point $t_j$ in turn, simulate $\Y_i(t_j)$ conditionally on $\X_i$ and previous simulated values $\{\Y_i(t_{k: k<j})\}$.   The observed data informing this is either longitudinal data (Section~\ref{sec:long}) or repeated cross-sectional data describing different times (Section~\ref{sec:long_csd}).  Simulating sequentially in time allows alternative risk-modifying interventions to be applied if necessary, though in this paper we consider only simulating under a ``status quo'' scenario matching the observed data.
\end{itemize}

\noindent Each time step of the simulation involves sampling from the multivariate distribution of $\Y(t)$.  Broadly, this is achieved either by:
\begin{itemize}
\item fitting a multivariate parametric model to the data, then simulating from the fitted model (parametrically)
\item drawing observations directly from the data (non-parametrically)
\end{itemize}
Table~\ref{tab:summary} summarises the available methods, categorising the broad principles underlying them, and choices to be made when designing them.  In Sections~\ref{sec:cross}--\ref{sec:long_csd}, we describe them in detail. 

\newcolumntype{P}[1]{>{\raggedright\arraybackslash}p{#1}}
\pretolerance=10000
\begin{table}
\raggedright
  \begin{tabular}{|P{1.5in}|P{1.5in}|P{1.5in}|P{1.8in}|}
    \hline
    Core principle & Variant & \multicolumn{2}{|c|}{Choices} \\ 
    \hline
    \multicolumn{4}{|p{5.2in}|}{\textbf{(a) Simulating cross-sectional data at one time (e.g. baseline): multivariate regression procedures (Section 3)}}\\
    \hline 
    Parametric & Conditional univariate & Conditioning order, outcome distribution, covariates, variance model & \multirow{5}{1.4in}{Reweighting to demographic data}\\
                   & Direct multivariate & Outcome distribution, covariates, variance model & \\
    \cline{1-3}
    Nonparametric resampling & Row-wise &  & \\
                              & Ad hoc variable-wise & Strata choice & \\
                              & Decision tree & Tuning  & \\
    \hline 
    \multicolumn{4}{|l|}{\textbf{(b) Simulating longitudinal data given longitudinal data (Section 4)}}\\
    \hline 
    Parametric & \multirow{2}{*}{Markov} & \multirow{2}{*}{Markov order} & \multirow{3}{1.4in}{Choice of multivariate regression procedure (block (a))} \\
    \cline{1-1}
    Nonparametric resampling &  &  & \\
    \cline{1-3}
    Parametric & Random effects &  & \\
    \cline{1-3}
    \hline 
    \multicolumn{4}{|l|}{\textbf{(c) Simulating longitudinal data given repeated cross-sectional data (Section 5)}}\\
    \hline 
    Rank/quantile stability & Quantile matching & What to match on, stratification &  Choice of cross-sectional multivariate regression procedure (block (a)) \\
                            & Shortest distance & Stratification & \\ 
    \hline
  \end{tabular}
  \caption{Summary of methods for longitudinal microsimulation of multiple chronic disease risk factors, with examples of choices involved in designing them.}
  \label{tab:summary}
\end{table}
\pretolerance=100

\section{Cross-sectional microsimulation} 
\label{sec:cross}
Both steps (a) and (b) in Section~\ref{sec:overview} involve
generating synthetic cross-sectional data at a particular time, given
data at that time.  Here we describe step (a): how the baseline
population is simulated.  We suppose that demographic characteristics
of the population are known from national administrative data such as
censuses, published as cross-tabulations (e.g. counts of people in
categories defined by, e.g. age group, sex and ethnicity), while both
demographic and health-related data are available from
individual-level surveys.  In some contexts, the goal is to simulate
a population of a specific area in geographical detail, known as
\emph{spatial}
microsimulation~\citep[e.g.][]{spatial_microsim,spatial-obesity,smoking_tomintz}.
By contrast, our focus is on simulating \emph{longitudinal} data. 
Therefore, in this section, we focus on describing techniques that serve
as the foundation for the methods used to simulate longitudinal data
(Sections~\ref{sec:long},\ref{sec:long_csd}) --- essentially, these are methods
for regression with a multivariate outcome.

\subsection{Nonparametric resampling from data}
\label{sec:cross:nonpar}

If survey data are comprehensive and representative enough, the simplest way to generate a synthetic population is to sample with replacement from those data, using, if necessary, sampling weights provided with the data~\citep[e.g.][]{lifesim, pandya_cvd, fem_goldman, archer, su_obesity, an_covid}.   Alternatively, if the administrative data is thought to better represent the population of interest to the policy question, we could define a synthetic population with the same proportions $p_s$ in category $s$ as in the published cross-tabulations~\citep{may_cvd,singh,fem_japan,mytton} .  The health-related data are then sampled from the subset of the survey data in category $s$, or from published models developed from survey data~\citep[e.g.][]{lim}.

Consider in more detail the problem of sampling $(\Y | \X=s)$ from
individual data with $\X=s$ (supposing, for this section, that $\X$
comprises the variables obtained from the census, while $\Y$ represents
everything else).  This is commonly done ``one row at a time'', so
that the synthetic vector $\Y_i$ can only take the \emph{combinations
  of values} that are observed from individuals in the data.  This
accounts for correlation between different variables, but a potential
disadvantage is that if the subset with $\X=s$ is small, then the
resulting distribution of $(\Y|\X=s)$ will be coarsely concentrated on
only a few values, which limits the ability to describe subgroups of
the population.  Missing data would also be awkward.

An alternative is to sample ``one variable at a time'', as done in
Mytton et al. \cite{mytton} and Kingston et al. \cite{kingston}, by sampling $\Y_{i,r} | \X = s$ from
observations of the $r$th variable, independently of other variables. 
However, this does not represent correlation between the variables.

\subsubsection{Stratified conditional sampling}
\label{sec:strat_cond}

A compromise between these alternatives is to do \emph{stratified}
sampling. For illustration, drop the conditioning on $\X$, and suppose
we want to sample from the multivariate distribution of $\Y$ in a
particular demographic subgroup.  The idea is to express this joint
distribution as a product of conditional distributions.  For example,
if there are only two variables, say, body mass index ($BMI$) and
total cholesterol ($TC$), the joint distribution $P(BMI,TC)$ can be
expressed as $P(TC)P(BMI|TC)$.  We can then sample from (an estimate
of) this distribution as follows.

\begin{itemize}
\item[(a)] sampling from $P(TC)$ by drawing a value ($y_1$, say) from
  the observed values of total cholesterol,

\item[(b)] sampling from $P(BMI|TC)$ by binning the data into strata
  defined by ranges of $TC$, and taking the $BMI$ value from a
  randomly-chosen individual in the stratum that contains $y_1$.
\end{itemize}
To generalise this to more than two variables, the multivariate
distribution can be decomposed into, say, a product of univariate
distributions,
\begin{equation}
  \label{eq:mv:cond}
  P(\Y)  =  P(Y_1)P(Y_2|Y_1)P(Y_3|Y_2,Y_1)...P(Y_R | Y_1,\ldots,Y_{R-1}) 
\end{equation}
or a product of simpler multivariate distributions.  We start by

\begin{itemize}
\item[(a)] sampling $Y_1$ from the full observed data. 

\item[(b)] Then for $r>1$, to sample $P(Y_r | Y_1,\ldots,Y_{r-1})$,
  the data are binned into strata defined by one or more of the
  variables that comprise $(Y_1,\ldots,Y_{r-1})$.  $Y_r$ is drawn
  from the observed values from people in the stratum that contains
  the values we have already drawn for $Y_1,\ldots,Y_{r-1}$.
\end{itemize}

This method relies on choosing which variables to stratify by, and how
to discretise or categorise each one of them.  The only use of this
idea in health microsimulation that we know of was by Mytton et al. \cite{mytton}
(in its generalisation to longitudinal data, see our
Section~\ref{sec:long}) who stratified according to \emph{ad hoc}
background judgements of which variables are predictors of other
variables, and initially chose a large number of categories. Then
during the sampling procedure, the categories were merged at stage (b)
if necessary, to ensure there is at least one individual in the
observed data in the corresponding stratum.  This strategy was
intended to describe the data in the greatest detail possible.

\subsubsection{Classification and regression trees}
\label{sec:cart}

We observe that the stratified sampling procedure is an example of the
machine learning techniques known as \emph{classification and
  regression trees} or \emph{decision trees}, which gives a more
formal approach to choosing the strata.  A brief outline is given
here, as fuller descriptions are given in many other
sources~\citep[e.g.][]{elements}.  These methods partition the space
of predictors in the style of a tree.  Given a starting partition
(e.g. men versus women) further ``branches'' are defined by splitting
each subspace by additional variables (e.g. BMI $>=26$ or $<26$).  The
tree is ``grown'' by adding sub-branches based on further splitting
conditions.  Given the final tree, a prediction for a given set of
predictor values can be made by finding the ``terminating'' branch, or
subset of the data, containing those values. The outcome variable can
then be predicted as the average of the outcome values within that
subset.  The choice of variables to split by, and how to discretise
them, is made by optimising a goodness-of-fit criterion such as mean
squared error, typically as part of a cross-validation process.
Software is available~\citep[e.g.][]{rpart} to implement variants of
this approach, and extensions such as random
forests~\citep{randomforest}.

This provides a statistically-principled approach to choosing the
strata required to estimate $P(Y_r | Y_1,\ldots,Y_{r-1})$ in
Section~\ref{sec:strat_cond}.  Synthetic data can be sampled by
drawing a random outcome value from the terminating subset.

\subsection{Parametric modelling of the joint distribution: conditional univariate regression models}
\label{sec:param:cross}

An alternative route to sampling from the joint distribution of risk factors $P(\Y)$ is to fit a fully parametric model to the data~\citep[e.g.][]{kypridemos,kypridemos_salt,kingston,breeze2016,breeze2020}.  A particularly easy way is to decompose the multivariate distribution as a sequence of univariate distributions, as in Equation~(\ref{eq:mv:cond}).  Each $P(Y_r | Y_1,\ldots,Y_{r-1})$ can be estimated in standard software by a regression (e.g. linear or logistic) of the outcome $Y_r$ on predictors $Y_1,\ldots,Y_{r-1}$.  Synthetic data can then be generated by sampling from the fitted models in order: 

\begin{enumerate}
\item sample from the model for $P(Y_1)$, producing simulated data $Y_1=y_1$, say.
\item sample from the second model $P(Y_2 | Y_1=y_1)$ using the simulated value in step 1 as the predictor.
\item continue sampling from $P(Y_r | Y_1,\ldots,Y_{r-1})$ for each $r$, using the previously simulated values as the predictors.
\end{enumerate} 
Instead of sampling directly from the parametric model, Alfons et al.~\cite{alfons} estimated mean risk factors from a parametric model, and then synthetic data were generated by sampling from the residuals and adding to the mean.  Alternatively, instead of decomposing into conditional univariate distributions, a multivariate distribution (conventionally a normal) could be fitted directly to data transformed to a suitable scale.  Discrete variables might be modelled by discretising a latent normally-distributed quantity, or modelled separately \citep[e.g.][]{fitzmaurice-laird}.  Similar multivariate modelling techniques are used in the context of multiple imputation of missing data~\citep[see, e.g.][]{huque2018comparison,cao2022review} for reviews.

For health microsimulation, we do not see any particular advantage in these alternatives, given the ease of use of the conditional univariate approach.  However, any models, and choices within those models (such as selection of predictors, or the order of conditioning) can be compared in terms of goodness of fit to the data, or informed by background information about causal relationships, say, between demographic, biological and behavioural variables~\citep[see, e.g.][]{kypridemos}.

\section{Longitudinal microsimulation using longitudinal data}
\label{sec:long}

The framework of multivariate modelling of a joint distribution
$P(\Y)$, set up in Section~\ref{sec:cross}, can be extended to perform
step (b) from Section~\ref{sec:overview}, where synthetic data on
multiple risk factors are simulated through time.  This section
details how this can be done with observed longitudinal data.

Statistically, the new requirement in this section is to express
correlation between the measurements at different time points, say
$\Y_i(t_1), \Y_i(t_2)$, from the same individual $i$.  This is the
domain of longitudinal data analysis, though for microsimulation, we
have the specialised requirement of modelling a multivariate outcome.
The key technique that we focus on is \emph{Markov} or
\emph{transition} modelling, which is naturally suited to longitudinal
microsimulation and has been used extensively for this
\citep[e.g.][]{kopec,markov-ecig, kingston, may_cvd, kasajima_cvd, richardson2022microsimulation, quan2023potential, heun2025dynamic}.

Section~\ref{sec:cross} described multivariate
cross-sectional regression models for a particular time $(t=0)$. To
extend these, we repeat the model at later times, while introducing
the \emph{outcomes from previous times} as predictors.  Thus no extra
tools are required.  We concentrate here on the first-order Markov
model, which uses only the most recent previous outcome. This
describes the joint distribution of risk factor variables at time
$t_j$ as
\begin{equation}
  \label{eq:markov}
  P\big(\Y(t_j) | \X, \Y(t_{j-1})\big)
\end{equation}
Synthetic data at time $t=0$ are simulated using one of the methods in
Section~\ref{sec:cross}.  Then for each subsequent time, a model of the
form~(\ref{eq:markov}) is fitted, conditioning on previously-simulated
data.  Synthetic data are simulated from the fitted model, and used as
the predictor for the next time.

\subsection{Conditional univariate Markov regression models}
\label{sec:markov}

As shown in Section~\ref{sec:param:cross}, a particularly
straightforward way to implement the time-specific model is via a
sequence of conditional univariate models.  Previously, we modelled the
$r$th risk factor variable at time 0 in terms of the
previously-modelled variables, $P(Y_r | Y_1,\ldots,Y_{r-1})$.  This is
extended to apply to longitudinal data at time $t_j$, by adding the
$r$th variable \emph{at the previous time} $t_{j-1}$ as an extra predictor:

\[
  P\big(Y_r(t_j) | Y_1(t_j),\ldots,Y_{r-1}(t_j),   Y_{r}(t_{j-1})\big)
\] 

\noindent Thus we express both correlation between different variables
at the same time, and correlation over time in the same variable $r$.
Higher-order Markov models could also be used with no greater
difficulty if appropriate data are available
\citep{wolfson-healthpaths,kopec}.  Note also that we could develop
prediction models for each risk factor
\emph{independently}\cite{hoerger2023new}, though this would not
account for any correlation \emph{between} different risk factors in the
way they change over time.

Any of the cross-sectional models described in Section~\ref{sec:cross}
can be extended in this way.  This includes the stratified sampling
and tree-based methods --- the previous response is simply considered
as an additional predictor.  For example, continuous outcomes could be
handled by linear regression, and binary outcomes or events by
logistic regression or time-to-event modelling.  Recurrent events or
counts could also be treated as binary (e.g. the risk of becoming
pregnant in the next time step, which may depend on the number of
previous pregnancies).

In practice, this idea is implemented by rearranging a longitudinal
dataset into ``transition'' format: with one row per \emph{unit time
  interval}, rather than one row per observation.  The observation at
the start of the interval can then be treated as a ``covariate'' in
regression modelling software, and the observation at the end of the
interval is the response.  Once the models are fitted, synthetic data
can be simulated for each time $t_j$, starting with the synthetic
population at time $0$, by using the \emph{simulated} response at
$t_{j-1}$ as a predictor.

\subsection{Random effects modelling} 
\label{sec:randomeffects}
An alternative to Markov modelling for representing multi-outcome
longitudinal data is based on random effects.  This was used by
Breeze et al. \cite{breeze2015,breeze2016,breeze2020} with a specific class of
random effects model described as a ``latent growth'' model.  In this
approach, a parametric function for the expected underlying trajectory
of a risk factor over time for an individual is specified, and
correlations between multiple outcomes for the same individual at the
same time point are expressed through shared random effects.  This
approach requires more specialist software than the Markov conditional
regression approach, which can just be built from coupling standard
linear or logistic regressions.  Again, if necessary, models based on
different approaches can be compared in terms of fit to a particular
dataset (Section~\ref{sec:checking}).

\subsection{Summarised longitudinal incidence data}

For binary variables $r$ for which we are interested in the incidence
over time, it may be sufficient to use published incidence rates or
probabilities to inform $P\big(\y_r(t_j) | \X, \Y(t_{j-1})\big)$,
rather than an individual longitudinal dataset, if these are available
by all relevant predictors.  Since this approach has generally been
used for variables considered as outcomes, rather than disease risk
factors, it is discussed in Section~\ref{sec:outcomes}.

\section{Longitudinal microsimulation using repeated cross-sectional data}
\label{sec:long_csd}

\begin{figure}
  \includegraphics[width=0.5\linewidth]{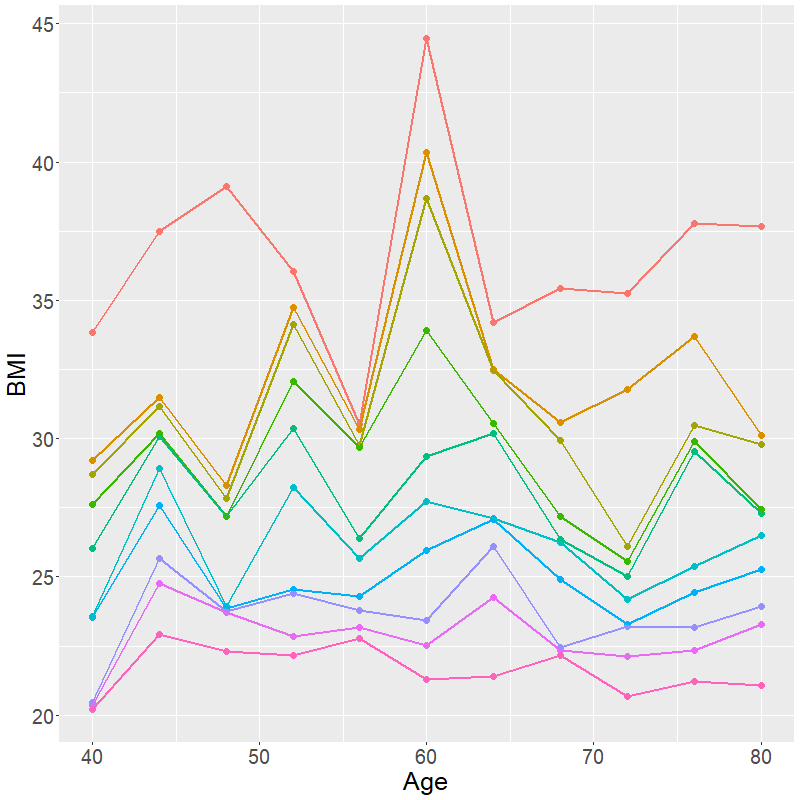}
  \caption{Generating synthetic longitudinal data from repeated cross-sectional data by rank stability.  Baseline cross-sectional data for BMI at age 40 are generated.  At age 44, a further cross-sectional dataset is generated.  The observations at age 44 are then connected to those with the same rank among the previous age-specific dataset.  The process is repeated at steps of 4 years of age to produce a set of synthetic individual trajectories.}
  \label{fig:rs}
\end{figure}

Suppose no longitudinal data are available, but instead we have
cross-sectional data that covers a range of ages, or different
calendar times.  This can still be used to simulate synthetic
longitudinal data where individual measurements are correlated through
time.  The key assumption generally used to do this is \emph{rank
  stability}.  This is is the assumption that an individual's rank or
quantile among other individuals, for a given variable, stays constant
over time \citep{lim, basu, kypridemos, rankstab_brazil}.  A basic
outline is as follows, and is illustrated in Figure~\ref{fig:rs}.

\begin{enumerate}

\item Data $(\y_i(0),\X_i)$ are generated for a set of $I$ synthetic
  individuals $i$, from a model for cross-sectional data at the baseline
  time $t_0=0$, as in Section~\ref{sec:cross}.
  
\item For each individual $i$ from step (1), the quantile of their
  simulated value $y_{ir}$ of risk factor $r$ is estimated (i.e. the value
  $p$ such that $P(Y_r < y_{ir}) = p$) from the baseline data.

  
\item A model for cross-sectional data at the next time $t_1$ is developed, and
  used to generate values for this risk factor for a similar
  population at that time.

\item Individual $i$'s risk factor value at $t_1$ is then set to the
  value at quantile $p$ among the population generated in step (3).

\item This process is repeated for all risk factors $r$ thought to be
  serially correlated.  Any variables assumed to not be serially
  correlated can be assigned at random from the data generated in (3),
  rather than quantile-matched.


\item The process is repeated over all times $t$, to produce a
  synthetic longitudinal dataset.
\end{enumerate}

For each risk factor, the resulting synthetic longitudinal dataset
will have the same cross-sectional distribution as the data at each
time, provided that the model in step (3) is well-fitting.  For
example, such a dataset might represent the increase in both the mean
and variance of cholesterol as a population ages.  In addition, people
with higher cholesterol than others at baseline will continue to have
higher cholesterol than others as they grow older.

However, the \emph{joint distribution} of risk factors at each time
will not necessarily be reflected by this procedure.  Any assumptions
made about these correlations in the cross-sectional model in step (3)
will be overridden, since (in step 5) trajectories of different risk
factors are produced independently.  To account for these correlations
explicitly, the procedure could be modified in the following alternative
ways:

\begin{itemize}
\item[(a)] \textbf{Stratified rank stability.} In parts (2,4), instead of the full population, the
  quantile is calculated with reference to a \emph{stratum} of the
  population defined by some time-constant variables (such as
  ethnicity or sex) \citep{kypridemos}.  This accounts for any
  correlation between the time-varying risk factor and these
  variables.

\item[(b)] \textbf{Matching on a variable of interest.} In part (5),
  instead of generating trajectories for different risk factors $j$
  independently, they could be generated jointly.  For example, we
  could match on the basis of one particular variable of interest that
  is thought to be rank-stable over time, e.g. an overall risk score
  \citep[such as QRisk,][]{qrisk}.  We would then assign \emph{all}
  time-varying risk factors for individual $i$ from the individual
  simulated at step (3) with the matching quantile of the risk score.
  If the risk factors are indeed correlated with the risk score, this
  gives an efficient procedure for generating them all simultaneously.

\item[(c)] \textbf{Shortest distance.} The method of Suen et
  al.\cite{suen} is based on jointly matching \emph{multiple}
  variables simultaneously between two time points to minimise the
  total amount by which those variables change over time, though this
  is more computationally intensive than procedures (a) and (b).  With
  one variable, this will simulate the same data as the basic rank
  stability procedure (1--6 above).  See Church~\cite{church:phd} for
  more details.
\end{itemize}

The choices involved in such approaches (e.g. whether to stratify,
choice of cross-sectional model, how to quantile-match) are often
bias-variance trade-offs: more detailed models may be less biased, but
give noisy predictions with smaller samples.  Background information
might sometimes be available to inform the choice.  In any case, the
simulated population distributions at each time can be compared
against the original cross-sectional data.  However, the rank/quantile
stability assumption cannot be checked directly, due to the lack of
longitudinal data.  In Section~\ref{sec:app:deident} we investigate
this assumption by removing individual identifiers from a real
longitudinal dataset.

While the rank stability procedure was designed for continuous variables, in principle, it could also be used (coarsely) for binary or ordinal discrete variables, though we are unaware of this being done.  In chronic disease microsimulation, it is more common to simulate binary variables using a longitudinal data source on incidence or risk (Sections 4.3 and 7, e.g. diabetes diagnosis).

\section{Empirical procedures to check and compare models}
\label{sec:checking}

If the microsimulation procedure is based on parametric models fitted
to observed data, these can be compared through standard likelihood
theory.  For example, Akaike's information criterion (AIC) gives a
convenient generic measure of predictive
ability~\citep{burnham-anderson}.  However, we want to compare between
parametric and nonparametric procedures for microsimulation, and to
show how well they predict specific outcomes of interest.  We describe
two broad approaches.

\subsection{Informal checks of fit to data}

Most simply, we can compare the synthetic data generated from the
microsimulation procedure with the observed data used to construct it.
For procedures to generate \emph{cross-sectional} data
(Section~\ref{sec:cross}), we could compare summaries of the marginal
distribution of particular variables (see
e.g. Figure~\ref{fig:cross:dens}), or of the joint distribution of two
or more variables if their correlations are of interest.

For procedures to generate \emph{longitudinal} data, there are two
aspects of microsimulations that should be checked:
\begin{itemize}
\item[(a)] individual-level trajectories: how successive observations
  from the same person are related.  This is only checkable against
  observed data if we have built the model from longitudinal data
  (Section~\ref{sec:long}, not~\ref{sec:long_csd}).
\item[(b)] population-level distributions: how distributions of risk
  factors change over time.  This can be simply checked against the
  cross-sectional (or slice of longitudinal) data at each time.
\end{itemize}

These considerations roughly correspond to two desirable
characteristics of a microsimulation model: its ability to represent
the mechanisms of risk factor evolution (hence to form a vehicle for
expressing the causal effects of interventions that modify risk
factors for some individuals) and its ability to describe the
progress of a population over time.  In each case, a simple plot, as
in the examples in Section~\ref{sec:app:long},
(Figures~\ref{fig:long:traj}--\ref{fig:long:agedists}) can be
sufficient to detect areas of potential lack of fit.

\subsection{Cross-validatory assessment of predictive ability}
\label{sec:cv}

A more formal approach is needed if we have to choose between a set of
models which all appear to fit reasonably when simulations are
informally compared with the data.  In general, cross-validation could
be used, a standard technique to assess the ability of a model to
predict left-out observations.  This enables parametric models to be
compared with nonparametric or ``machine learning'' procedures, and
is related in principle to AIC~\citep{stone1977asymptotic}.

Cross-validation requires the choice of a target of prediction and a
measure of accuracy.  Accuracy can be measured by root mean squared
error (RMSE) for continuous variables, or classification error for
discrete variables.  For procedures to generate cross-sectional data
(Section~\ref{sec:cross}), we would simply predict the variable or
variables of interest, either a specific disease risk factor or an
overall risk score, depending on what is most important to the health
policy question.  For the procedures in Section~\ref{sec:long} that
generate longitudinal data sequentially through time, it is natural to
assess how well a model predicts future values of risk factors, most
simply, one step ahead in time.  This is limited by what observed data
are available --- for example in our case study data (ELSA,
Section~\ref{sec:casestudy}), there are only two observations per
person, four years apart, so a longer-term assessment would not be
possible.

Judging \emph{individual-level} predictive ability is not necessarily
a good guide to a model's ability to simulate population data,
however.  For example, if using models of the form in
Section~\ref{sec:long} for \emph{individual} risk prediction, we would
generally base the risk estimate on \emph{expected} values of risk
factors $E(Y)$ (\emph{deterministic} predictions).  By contrast, a
microsimulation model aims to generate synthetic data $Y_i$, which are
more realistic if simulated \emph{stochastically} from the model.
However if performance is quantified solely by mean squared error of
an individual prediction, then a stochastic simulation would be judged
as worse, as it adds noise to a deterministic prediction.

We would recommend a combination of the two approaches --- using plots
of long-term predictions to check that the microsimulation procedure
generates believable data over the long term, while using formal
short-term assessments if needed for fine tuning and selection of
statistical models.

\section{Simulating outcomes given risk factors}
\label{sec:outcomes}
  
In longitudinal microsimulation models in health, two broad approaches have been taken to determine the probability that a simulated person $i$ gets a disease or other outcome at a time $t_{j+1}$, given risk factors $y_i(t_j)$ (in the notation of Section~\ref{sec:overview}).  Here we briefly review these, and how these are related to the choice of model for simulating risk factors.

In one approach, the outcome probabilities are estimated from the same \emph{individual-level} longitudinal data source used to infer risk factor progression.  Effectively, outcomes are treated in the same way as risk factors, as an additional variable $r$ in the set of $R$ variables simulated.  This has been done when the dataset covers the time period and population of interest, and informs all effects of predictors on the outcome.  For example Kingston et al.\citep{kingston:multimorb,kingston} modelled diabetes, cognitive impairment and several other conditions in this way as part of a model to forecast multimorbidity and long-term care needs of the older population in the UK, using cohort data for this population, while May et al. \cite{may_cvd} used a similar approach in the context of Ireland.  In this approach, the statistical model is chosen according to the same considerations as the models in Section~\ref{sec:long} --- essentially we want the model to fit the outcomes and risk factors of interest (Section~\ref{sec:checking}).

This can be contrasted with approaches where a \emph{summarised} (non-individual) data source is used to inform outcomes.    For many disease outcomes, there are published risk prediction models, built from cohort data.  These give an estimate of the risk of experiencing an event in some future period, given a set of risk factors.  For example, in the health checks model, QRisk~\cite{qrisk} and CAIDE~\cite{caide}, were used to predict myocardial infarction or stroke, and dementia, respectively.   Using published disease incidence rates is conceptually similar to using a risk prediction model, though such data are designed for describing populations or population subgroups rather than individualised risk.  This may be more appropriate if the model does not need to reflect the effect of several risk factors, and the published data describe the population of interest, or a detailed risk prediction equation is unavailable.  For example, in their model for forecasting osteoarthritis burden, Kopec et al~\cite{kopec} used summarised population-wide incidence data by age and sex, adjusted for BMI (the main modifiable risk factor) using hazard ratios estimated from a smaller longitudinal dataset. When using summarised data, ad hoc adjustments may be needed if the time period of the data is different from the time unit of the model, for example, Mytton et al.~\cite{mytton} converted from 10 year to 1 year risks while calibrating to achieve a smooth change in risk by year.

Again, whether risk scores or summarised incidences are used to model the outcome, the risk factor model should be chosen to represent the risk factors that matter to the modelling question.  If outcomes are generated by using a risk score, the model should represent that score well, as well as any particular risk factors that are targeted by interventions.   This principle is demonstrated for QRisk in the health checks case study in section~\ref{sec:casestudy}.

\section{Intervention effects: inference and simulation}
\label{sec:interventions}

Here we discuss how microsimulation models are designed to represent the effects of interventions, and how this relates to the choice of model to simulate risk factors.  In all the applications that we are aware of, the model describes at least a ``baseline'' or ``status quo'' scenario, where risk factors and outcomes are generated based on recent population data.  The purpose of the microsimulation is either to forecast the future if this baseline were to persist, or to predict changes in outcomes under alternative scenarios.  The model is structured to represent the causal effects of any risk-modifying interventions which characterise the alternative scenarios, and those effects are inferred from sources outside the ``baseline'' data.

For example, in the health checks model~\cite{mytton}, the baseline data were taken from before the health check programme was initiated.  The model was designed to assess the effect of the programme, and the potential effects of changes from current practice.   The effect of health checks is represented in the model through the effects of specific treatments started following a health check (e.g. statins that lower cholesterol), and rates of prescription and adherence for these treatments.  The effect of statins on cholesterol was obtained from a meta-analysis of randomised clinical trials, and other data directly inform prescription and adherence rates.  Thus there was no need for causal inference of the effect of health checks (or any of its components) from the baseline data.

Then to apply the intervention effect in the microsimulation model, risk factor trajectories are perturbed, using the available causal knowledge.  In a Markov transition model (Section~\ref{sec:markov}), this is straightforward --- for example, when statins are prescribed and taken following a health check, the cholesterol in the next time unit was lowered by the corresponding effect inferred from trials.  This assumes that the future of two otherwise-identical people with the same cholesterol level does not depend on whether that level was achieved with the aid of statins.   In the rank stability method (\ref{sec:long_csd}), after a risk factor is changed by an intervention, quantiles can simply be recalculated and assumed to persist from that point.   The random effects model (\ref{sec:randomeffects}), in which cholesterol would be expressed as a parametric function of time and an individual random effect, is less amenable to this kind of dynamic modification of the modelled variable. 

In summary, the microsimulation approach is based on explicitly describing the causal mechanism behind interventions, so assumes the full mechanism is known and effects can be easily quantified.  Hence there are limitations when this does not hold.  For example, the effect of statins on disease incidence observed in trials~\cite{fulcher2015efficacy} is more than would be expected from just cholesterol reduction --- an ad hoc approach to calibrate the effect of statins was used in Mytton et al. \cite{mytton}.

\section{Case study}
\label{sec:casestudy}

We demonstrate the use of these methods, and how they are compared, in
the context of a typical example.  We suppose the microsimulation
model is designed to assess the effect of a ``health check'' programme
that can lead to the reduction of one or more cardiovascular risk
factors, as in Mytton et al.~\cite{mytton} (described in the
Introduction).  The risk factors are combined into the QRisk3 score,
which is used to simulate cardiovascular disease incidence.  Here we
compare models for simulating the risk factors, the same ones
considered by Mytton et al..  We primarily favour models that
accurately represent the QRisk3 score, while also representing well
the risk factors that are modifiable by treatments that may be given
after health checks (cholesterol, blood pressure, smoking and BMI).

The cross-sectional data available include demographic information
from the 2011 UK census, and from the Health Survey for England
(HSE)~\citep{HSEguide} an annual cross-sectional dataset, from which
we include the years from 2009 to 2018, with 108,711 participants from
all ages.  Longitudinal data are available from the English
Longitudinal Study of Ageing (ELSA) \citep{ELSAdata}, which followed
up a sample of earlier HSE participants from 1998 to 2019.  The risk
factors recorded in the HSE and ELSA include age, sex, ethnicity (as
white / non-white), quintiles of household income, body mass index
(BMI), systolic and diastolic blood pressure (SBP,DBP), total and HDL
cholesterol, glycated haemoglobin (HbA1c), diabetes diagnosis and
smoking status.  Given these variables, the QRisk3 score can be
computed, assuming the less common risk factors included in this score
are absent.  While ELSA participants are followed up every two years,
these biological variables are only recorded at four-yearly intervals,
so when the ELSA data are rearranged in ``transition'' format
(Section~\ref{sec:long}) they are restricted to the subset describing
individual risk factor changes over four-year time steps.  Our sample
from ELSA included 4,637 individuals (mean age 60 years) for whom
complete data for our analysis was available at more than one time
point.

We now compare a selection of the most flexible and easily applicable
procedures for simulating synthetic risk factor trajectories discussed
in Sections~\ref{sec:cross}--\ref{sec:long_csd}.  Code used for this
case study is available at
\url{https://github.com/oliverc-147/microsim}.

\subsection{Simulating baseline cross-sectional data}
\label{sec:app:csd}

We suppose that an initial goal is to develop a model to simulate risk factors $(\Y,\X)$ for a synthetic baseline population of 40-44 year olds in England in the year 2011.  This is done by developing a multi-outcome regression model for $(\Y|\X)$, where $\X$ includes time-constant demographic variables (year of age, sex, ethnicity and income) and $\Y$ comprises the remaining risk factors (which will vary over time).  The model is fitted to the subset of the HSE participants aged 40--44 years (6811 people).   Then to generate a synthetic population representative of England, we can then simulate data from $P(\Y | \X = s)P(\X = s)$, estimating $P(\X = s)$ from the census data.

\begin{figure}
\centering
\includegraphics[width=0.5\linewidth]{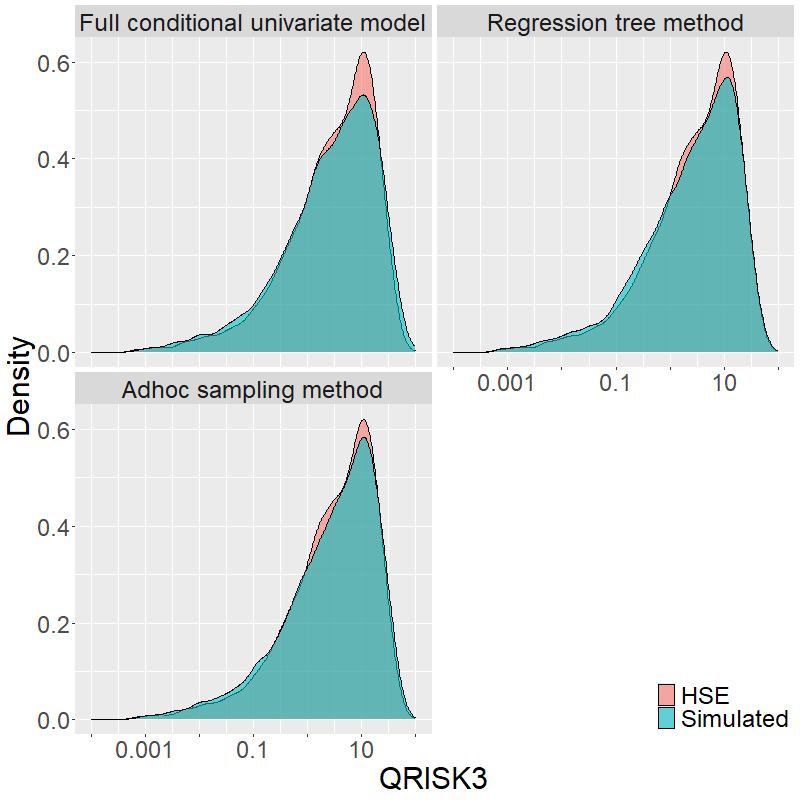}
  \caption{Comparison of kernel density estimates of the distribution of QRisk3 at baseline between simulated populations and the observed HSE data, under different simulation methods based on cross-sectional data.}
  \label{fig:cross:dens}
\end{figure}

We compare the parametric conditional univariate method (Section~\ref{sec:param:cross}, using standard linear, logistic or ordinal regressions without interactions) with two stratified sampling methods~(\ref{sec:cross:nonpar}), one using the same ad-hoc procedure as in \cite{mytton}, and one using a classification and regression tree, using the \texttt{rpart} package, tuned by cross-validation.  Details of how these procedures were implemented are given in the Appendix.   All these methods give satisfactory representations of the distribution of the QRisk3 score for HSE (Figure~\ref{fig:cross:dens}).   The choice between them could be made with the RMSE (from 10-fold cross-validation) of predicting a variable of primary interest, in this case QRisk3.  Table~\ref{tab:cross:rmse} indicates that the tree-based method gives the best prediction of QRisk3, though the optimal model might differ if different variables were of interest.

\begin{table}
\centering
\small
\begin{tabular}{l|lllllll}
& \multicolumn{7}{c}{\textbf{RMSE of variable}} \\
 & BMI & \multicolumn{2}{c}{Blood pressure} & \multicolumn{2}{c}{Cholesterol} & HBA1C & QRISK3 \\
 & & Systolic & Diastolic & HDL & Total & & \\
\hline
\begin{tabular}[c]{@{}l@{}}Full conditional \\univariate regression\end{tabular} & 7.07 & 21.4 & 15.0 & 0.57 & 1.60 & 10.6 & 6.38 \\
\begin{tabular}[c]{@{}l@{}}Ad hoc stratified\\ sampling \end{tabular} & 6.99 & 21.9 & 14.5 & 0.58 & 1.52 & 10.5 & 5.05 \\
\begin{tabular}[c]{@{}l@{}}Tree-based\\ sampling \end{tabular} & 7.04 & 22.0 & 14.6 & 0.58 & 1.50 & 10.6 & 4.90 \\
\end{tabular}
  \caption{Comparison of accuracy of regression models for cross-sectional data, in terms of root mean squared prediction error of particular variables.}
  \label{tab:cross:rmse}
\end{table}

\subsection{Simulating longitudinal trajectories given longitudinal data}
\label{sec:app:long}

Now we consider a situation where longitudinal data are available (as in \cite{mytton}).  We compare the following procedures to simulate synthetic longitudinal data for the same population as the baseline data, using the ELSA data.  Each of these is constructed as a Markov model, with previous values of risk factors treated as predictors, as in Section~\ref{sec:long}.
\begin{enumerate}
\item a fully parametric, conditional univariate regression model. Normal linear regression is used for continuous variables, and logistic or ordinal regressions for discrete variables, with linear effects of continuous predictors and no interactions.  Within this, we also compared the following alternatives:
  \begin{itemize}
  \item modelling the (log) \emph{variance} of the risk factors, as well as the mean, as a linear function of age, given evidence that systolic and diastolic blood pressure variance increase with age~\citep{bloodpressure_age2}.
  \item instead of including all recorded covariates, choosing covariates by backward stepwise selection on the basis of AIC.
  \end{itemize}
\item the stratified sampling method, using an \emph{ad hoc} choice of strata as in~\ref{sec:app:csd}.
\item a classification and regression tree, using the same tuning strategy as in~\ref{sec:app:csd}.
\end{enumerate} 

The models are used to generate simulated data for a baseline population aged 40-44, at four-yearly steps, for survivors up to age 100-104 (though note in a full health policy microsimulation model, some of these will be assumed to die earlier).

\begin{figure}
  \centering
\includegraphics[width=0.7\linewidth]{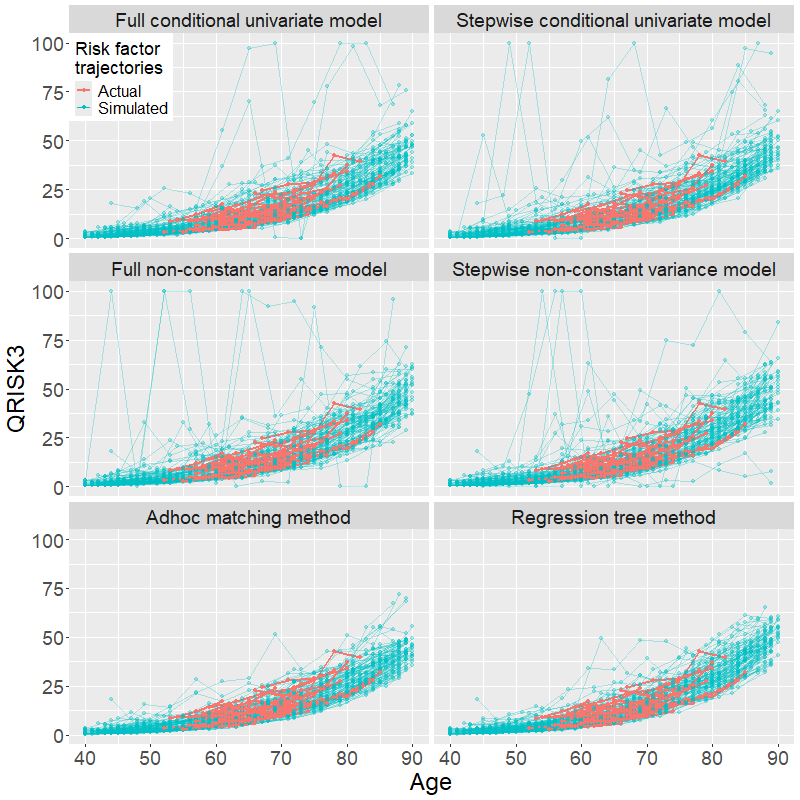}
  \caption{Trajectories of QRisk score simulated by alternative methods based on longitudinal data, compared with the observed ELSA data.}
  \label{fig:long:traj}
\end{figure}

\begin{figure}
  \centering
\includegraphics[width=0.7\linewidth]{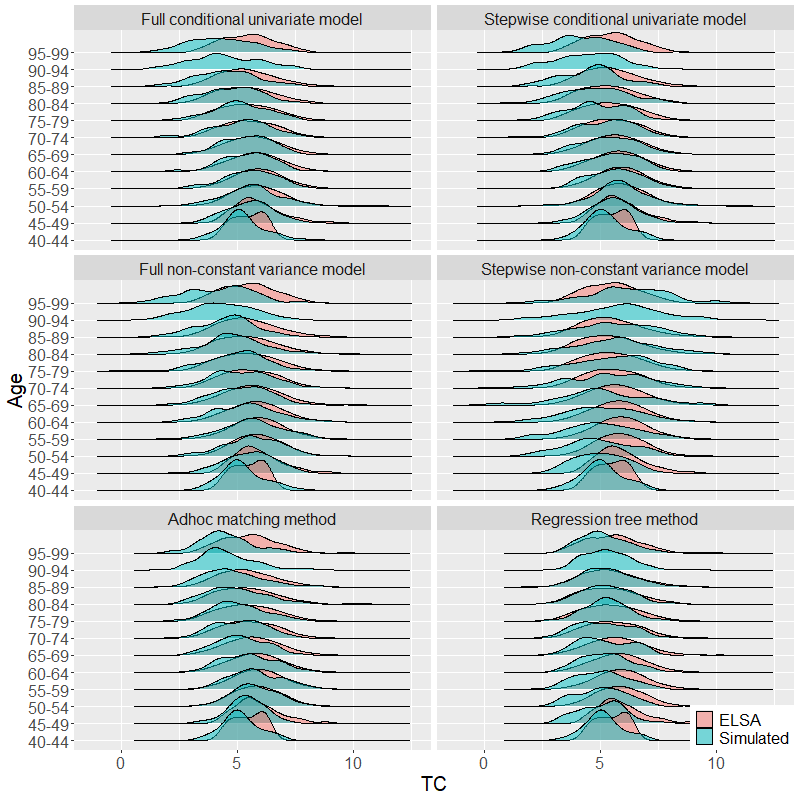}
\caption{Age group-specific distributions of total cholesterol from alternative simulation methods based on longitudinal data, compared with observed distributions from ELSA.  Kernel density estimates used.}
  \label{fig:long:agedists}
\end{figure}

The trajectories of QRisk score for a sample of synthetic individuals are illustrated in Figure~\ref{fig:long:traj}.  These simulated trajectories are overlaid over the original ELSA data (note the models enable modest extrapolation outside the range of ages represented by the data).  The models represent well the extent of individual-level variability over time observed in ELSA, except that the regression model built by stepwise selection appears to overestimate this variance for some individuals.  The simulated data from each model also reflects the change in the distribution of risk over the population as people age.  As an alternative illustration of this change, Figure~\ref{fig:long:agedists} compares the distribution of a specific risk factor (total cholesterol), for a series of age groups, between the simulated and observed data.

To select between models that give reasonable long-term predictions, we could again compare the RMSE of one-step predictions of variables of interest, as in Table~\ref{tab:long:rmse}.  In this case, the decision tree and the parametric model with non-constant variance and stepwise variable selection gave the best predictions of QRisk3.  For predicting specific risk factors, no other models were substantially better-performing than these.  Modelling outcomes on the log scale was also investigated, but this did not improve predictions.

\begin{table}

\centering
\small
\newcolumntype{P}[1]{>{\raggedright\arraybackslash}p{#1}}
\pretolerance=10000
\begin{tabular}{P{1.5in}|llllllll}
\multirow{2}{*}{} & \multicolumn{8}{c}{\textbf{RMSE of variable}} \\ 
\cline{2-9}
 & BMI & \multicolumn{2}{c}{Blood pressure} & \multicolumn{2}{c}{Cholesterol} & HBA1C & QRISK3 \\
 & & Systolic & Diastolic & HDL & Total & \\
\textbf{Using longitudinal data} &  &  &  &  &  &  &  \\ 
\hline
\textbf{\textbf{Parametric methods}} &  &  &  &  &  &  &  \\ 
\cline{1-1}
All covariates, constant variance & 2.96 & 20.4 & 12.1 & 0.35 & 1.29 & 6.34 & 4.63 \\
Stepwise selection, constant variance & 2.97 & 20.2 & 12.0 & 0.35 & 1.29 & 6.35 & 4.65 \\
All covariates, variance dependent on age & 2.94 & 20.5 & 12.0 & 0.35 & 1.27 & 6.31 & 4.68 \\
Stepwise selection, variance dependent on age & 2.98 & 20.4 & 12.1 & 0.35 & 1.27 & 6.36 & 4.62 \\ 
\cline{1-1}
\textbf{Non-parametric} & \multicolumn{1}{c}{} &  &  &  &  &  &  &  \\ 
\cline{1-1}
Ad hoc categories & 3.24 & 20.3 & 12.2 & 0.55 & 1.28 & 6.89 & 5.18 \\
Optimal decision tree & 2.98 & 20.3 & 12.2 & 0.36 & 1.29 & 6.35 & 4.41 \\ 
\cline{1-1}
\textbf{\textbf{Using cross-sectional data}} &  &  &  &  &  &  &  &  \\ 
\cline{1-1}
Rank stability; no strata & 2.20 & 16.6 & 9.85 & 0.27 & 1.05 & 4.88 & 4.33 \\
Rank stability; sex, income, ethnicity strata & 2.12 & 16.6 & 9.82 & 0.27 & 1.04 & 4.87 & 4.33 \\
Shortest distance; no strata & 2.56 & 16.8 & 10.0 & 0.29 & 1.07 & 5.13 & 4.58 \\
Shortest distance; sex, income, ethnicity strata & 2.69 & 16.7 & 10.1 & 0.30 & 1.07 & 5.06 & 4.58\\ 
\hline
\textbf{Mean of variable} & \multicolumn{1}{c}{27.9} & 132.9 & 74.5 & 1.60 & 5.68 & 39.8 & 14.4
\end{tabular}
\caption{Comparison of accuracy of Markov regression models for longitudinal data, in terms of root mean squared one-step prediction error of particular variables}
  \label{tab:long:rmse}
\end{table}
\pretolerance=100

\subsection{Simulating longitudinal trajectories given cross-sectional data}
\label{sec:app:long_csd}

As an alternative route to simulating synthetic longitudinal data, we
suppose that the longitudinal data from ELSA are not available, but
instead we use cross-sectional data, grouped by age, from the HSE.  We
compare the following implementations of the rank stability principle,
as explained in Section~\ref{sec:long_csd}:
\begin{itemize}
\item quantiles of every risk factor assumed stable over time, with each risk factor at a given time taken from a different individual in the data

\item matching all risk factors based on the quantiles of the QRisk score
\end{itemize}
In each of these, we also compare a stratified (by a combination of
sex, ethnicity and income quintiles) rank stability procedure with an
unstratified procedure.  The shortest distance method \cite{suen}
is also investigated.  In all implementations, the time-specific
cross-sectional data (steps (a) and (c) in Section~\ref{sec:long_csd})
are generated by from sampling individuals ``one row at a time'' from
the HSE.

\begin{figure}
  \centering
\includegraphics[width=0.7\linewidth]{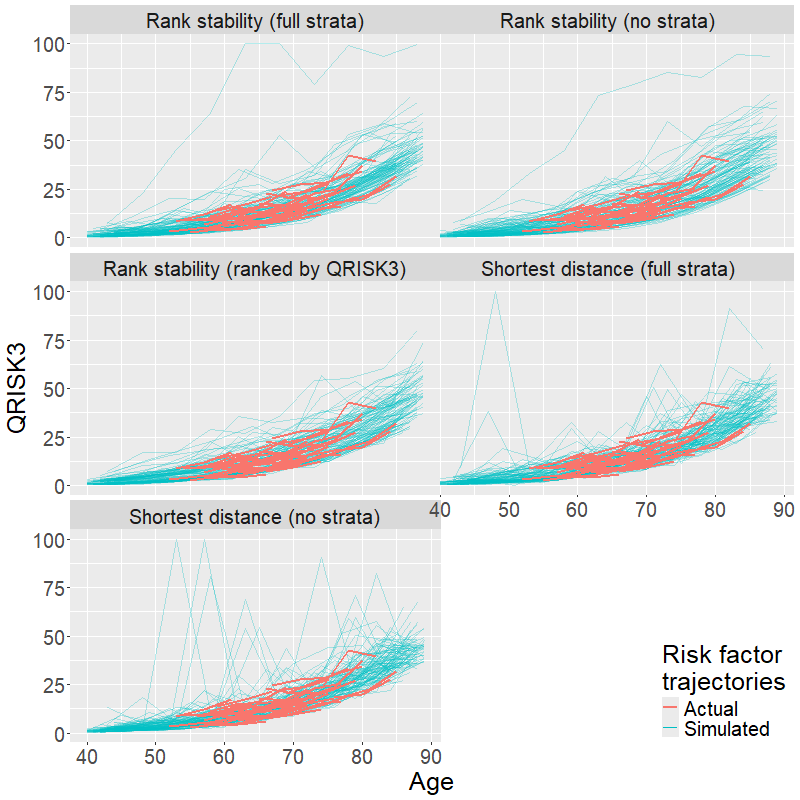}
  \caption{Synthetic trajectories of QRisk score generated by methods that use repeated cross-sectional data and different variants of the ``rank stability'' assumption. Full strata refers to stratification by sex, ethnicity and income groups.}
  \label{fig:traj:rank}
\end{figure}

Samples of the synthetic trajectories generated by these methods are
illustrated in Figure~\ref{fig:traj:rank}, alongside the cross-sectional data
that they were obtained from.  The population distribution at each time
is well matched by the cross-sectional data under each method, as expected.
A stricter test is how well the within-individual between-time
variability in the synthetic longitudinal data reflects that observed
in real longitudinal data --- but comparing with the real trajectories
from Figure~\ref{fig:long:traj}, they are qualitatively similar.

\subsubsection{Comparing rank stability methods using de-identified longitudinal data}
\label{sec:app:deident}

As a further investigation of the rank stability methods, we create
cross-sectional data from the longitudinal ELSA data by removing the
individual identifiers.  This is then used to simulate longitudinal
data, assuming rank stability.  The simulated data are then compared
with the original data, with the longitudinal identifiers restored.

More specifically, the longitudinal data $\{y_{ij}\}$ for individuals
$i$ and time points $j$ are rearranged into (start, end) pairs
$\{(y_{i1},y_{i2}), (y_{i2},y_{i3}),\ldots \}$.  representing individual
risk factor changes over a single (4-year) time step.  Then by
breaking the individual link in each pair, we create a cross-sectional
dataset with two cross-sections, comprising the ``start'' and ``end''
values respectively.  This dataset is used with the methods of
Section~\ref{sec:long_csd} to predict a second risk factor value for a
set of synthetic individuals with baseline risk factor values taken
from the ``start'' values.  Then by restoring the link, we can compare
the predicted value for each starting observation with the actual
corresponding ``end'' observation.  The overall accuracy of a
particular procedure is measured by the root mean squared error of
predicting a variable of interest.

Note this approach would not be used to choose between rank stability
methods in a \emph{specific} application, since rank stability would
probably only be used in situations where longitudinal data are
unavailable.  However, our investigation is intended to give some
evidence of the relative appropriateness of different rank stability
assumptions \emph{in general}.


We compare the relative accuracy of different variants of the rank
stability method (last block of Table~\ref{tab:long:rmse}).
Stratifying the rank stability procedure made little difference to the
predictive ability for any of the variables.  Matching each risk
factor independently gives better predictions of each specific risk
factor than matching them all using QRisk.  However, this does not
translate to better predictions of the QRisk score, for which matching
based on QRisk itself is better.  Which approach is preferred would
depend on which variable or variables are of interest to the policy
question.  If some (but not all) of the risk factors are of interest,
they might each be modelled by rank-matching independently, with the
remaining risk factors modelled by matching based on the overall risk
score.  The shortest distance method \cite{suen} gave reasonable
results in comparison, though note this is substantially more
computationally expensive than the standard rank stability method,
taking four times as long for matching among 1000 individuals, and
with computation time scaling in proportion to $n^3$, where $n$ is the
number of individuals in the stratum being matched
(see~\cite{church:phd} for full details).

Note with this ``de-identification'' procedure, we can also directly
compare the rank stability methods with the methods of
Section~\ref{sec:long} that use longitudinal data.  The estimates of
the RMSE of one-step predictions in Table~\ref{tab:long:rmse} are
based on the same reference data, so are comparable.  The rank
stability method based on matching on QRisk gives better short term
predictions than any of the methods that use longitudinal data.  In
the long term, however, realistic trajectories and population risk
distributions can be obtained using either longitudinal or repeated
cross-sectional data (Figures~\ref{fig:long:traj}
and~\ref{fig:traj:rank}).

\section{Discussion}

This paper has reviewed methods for modelling multivariate
longitudinal or repeated cross-sectional data for use in longitudinal
microsimulation models.  A range of useful and flexible methods are
available, either fully-parametric or nonparametric.  For the baseline
population (\ref{sec:cross}), stratified conditional sampling forms a
flexible framework for simulating multiple starting risk factors given
suitable demographic and health survey data.  For simulating
longitudinal progression of these variables, the first consideration
is what data are available that cover the population and time/age
range needed for the microsimulation.  Longitudinal data are ideal,
since they directly inform individual progression.  Longitudinal data
are not always available, however, and repeated cross-sectional data
are cheaper to collect and may be more likely to cover a wide range of
times and ages.  We have provided some evidence that
repeated cross-sectional data can serve as an acceptable substitute
for longitudinal data in microsimulation, by using a ``rank
stability'' assumption.

Within each of these model classes, the specific model to fit to data
can be chosen according to statistical fit.  Among parametric methods,
those that express multivariate distributions as a sequence of
conditional univariate distributions can be used with standard
software, and are capable of fitting a rich variety of data.
Classification and regression trees form a flexible nonparametric
alternative, which has available software.  Though in principle, any
predictive machine learning algorithm would also be applicable to this
problem (e.g.  neural networks, support vector machines).  For
longitudinal data, transition models are naturally suited to
simulation of data through time.  Extensions of methods previously
used in this area were proposed, e.g. modelling predictors of the
variance of a risk factor.  With repeated cross-sectional data, we
have elaborated the statistical principles behind the common ``rank
stability'' procedure, proposed some useful variants of it, and shown
its utility.  In our demonstration on a typical application to
cardiovascular risk factors, we showed how the best-fitting models can
be identified in practice.

The limitations of the modelling framework we have described include
the discrete-time assumption.  The time grids of the observed and
simulated data were assumed to be the same (in the CVD application
this was 4 years). This in practice limits the data to those that are
(or can be assumed to be) equally spaced.  To simulate data on finer
grids than the observed data, most simply we could interpolate, but
continuous-time modelling would be more flexible.  Random effects
models with explicitly continuous parametric functions of time could
accomplish this.

A potential weakness of some risk factor simulation models is
short-term variability in excess of that observed in real data (seen
in some panels of Figures 4 and 6).  In practice, if using
longitudinal data, values could be bounded by rejecting implausible
simulated values, or more formally by using Bayesian regression models
where plausibility is constrained through a prior.  With
cross-sectional data, the rank stability assumption appeared to
alleviate this problem.

While longitudinal data can represent changes in an individual,
changes in the population over time are often better described in
practice by repeated cross-sectional datasets.  For individual-level
simulations of a population health policy, both are important.
A useful avenue for further research might be to combine both
longitudinal and repeated cross-sectional data where these are both
available, for example via a joint likelihood, to make the
most of the strengths of both. 

The risk factor simulation models we investigated form only one aspect of typical longitudinal microsimulations for evaluation of policies or future health burden.  Aspects we did not investigate include how to model the effects of risk-modifying interventions that might distinguish different policies.  In the literature that we have reviewed, the causal mechanisms for any interventions being modelled are assumed to be known and estimates of the causal effects are readily available (Section~\ref{sec:interventions}).  In particular we have not considered situations where the ``baseline'' population data used as the basis of simulating risk factors is also used to inform the effects of interventions, which may be challenging to disentangle.  We also not did not investigate modelling disease outcomes in depth (Section~\ref{sec:outcomes}).  There may be challenges when the baseline data does not contain sufficient information about the outcome (e.g. if it is uncommon, or follow-up is too short), and external data on outcome incidence are coarsely stratified by risk factors or describe a different population.

In summary, this paper improves the knowledge base to support the use of longitudinal microsimulation models to inform health policy.

\subsection*{Acknowledgements}
The authors were supported by the Medical Research Council, programme code MC\_UU\_00040/4.   Many thanks to the editor, associate editor and two reviewers for their detailed scrutiny which helped to improve the paper.

\bibliography{bib1}

\appendix

\subsection*{Tuning of the classification and regression tree}

Classification and regression trees (CART) can be fit using the CART algorithm, the details of which are explained in \cite{elements}.  In our application, we fit regression trees using the \texttt{rpart} R package, with different combinations of values for the complexity parameter and minimum terminal node size. The complexity parameter regularises the model by penalising large tree sizes. The minimum terminal node size is the minimum number of observations that each of the final nodes of the decision tree should contain, meaning that we can fit decision trees with a large number of nodes, whilst ensuring that every node contains at least one observation. We focused on these two hyperparameters as they had the greatest effect on the fit of the model. 

We used a range of values for the from $10^{-5}$ to $1$ for the complexity criterion, and either 1 or 50 for the minimum terminal node size.  The results in Table~\ref{tab:cross:rmse} are based on the model that used a minimum terminal node size of 1 and a complexity criterion of $10^{-4}$, as this model exhibited the lowest RMSE for all outcomes.  The results in Table~\ref{tab:long:rmse} are based on the model that used a minimum terminal node size of 1 and a complexity criterion of $10^{-3}$.

\subsection*{Ad-hoc procedure for stratified conditional sampling}

The categories used in the ad-hoc sampling method were based those chosen in \cite{mytton}, to give as detailed a stratification as was practicable given the data.  Continuous variables are split into ranges according to their quantiles estimated from the HSE data.  Supplementary Table~\ref{tab:cs_match_table} details the categories used when simulating cross-sectional data.  When simulating longitudinal data, previous values of a given variable can also be used as predictors. Thus, the categories chosen when simulating longitudinal data, which are shown in Supplementary Table~\ref{tab:match_table}, are slightly different.

\subsection*{Supplementary tables}

\begin{table}[h]
\small
\begin{tabular}{|l|l|l|l|} 
\hline
\begin{tabular}[c]{@{}l@{}}Variable to \\ be simulated\end{tabular} & \multicolumn{3}{l|}{Predictor variables defining bins (with number of categories)} \\ 
\hline
 & First strata & Second strata & Third strata \\ 
\hline
BMI & Age (8), gender (2) & Age (4) &  \\ 
\hline
SBP & Age (3), BMI (3) & Age (3) &  \\ 
\hline
DBP & Age (3), BMI (3), SBP (11) & Age (3), BMI (3) &  Age (3) \\ 
\hline
TC and HDL & Age (3), gender (2), BMI (3) & Gender (2), BMI (3) & Gender (2) \\ 
\hline
HbA1c & Age (3), gender (2), BMI (3) & BMI (3) & \\ 
\hline
\begin{tabular}[c]{@{}l@{}}Smoking \\status\end{tabular} & Gender (2), age (4), BMI (3) & Gender (2), age (4) & Gender (2)\\
\hline
Diabetes & \begin{tabular}[c]{@{}l@{}}Age (3), gender (2), \\ BMI (3), HbA1c (6)\end{tabular} & BMI (3), HbA1C (6) & HbA1C (6) \\ 
\hline
\end{tabular}
\caption{Strata used to categorise synthetic individuals and observed individuals in sampling in the ad hoc stratified sampling procedure, with the second and third categories used if a match can't be found from the first categories}
\label{tab:cs_match_table}
\end{table}

\begin{table}[h]
\small
\begin{tabular}{|l|l|l|l|} 
\hline
\begin{tabular}[c]{@{}l@{}}Variable to \\ be simulated\end{tabular} & \multicolumn{3}{l|}{Predictor variables defining bins (with number of categories)} \\ 
\hline
 & First match & Second match & Third match \\ 
\hline
BMI & \begin{tabular}[c]{@{}l@{}}Age (8), gender (2), \\ smoking (5), BMI (12)\end{tabular} & \begin{tabular}[c]{@{}l@{}}Age (4), smoking (5), \\ BMI (3)\end{tabular} &  \\ 
\hline
SBP and DBP & \begin{tabular}[c]{@{}l@{}}Age (3), BMI (3),\\ SBP (11), DBP (6)\end{tabular} & \begin{tabular}[c]{@{}l@{}}Age (3), BMI (3), \\ SBP (11)\end{tabular} &  \\ 
\hline
TC and HDL & \begin{tabular}[c]{@{}l@{}}Age (3), gender (2), \\ BMI (3), EQV5 (5), TC (8)\end{tabular} & \begin{tabular}[c]{@{}l@{}}Gender (2), \\ TC (8), BMI (3)\end{tabular} & \begin{tabular}[c]{@{}l@{}}Gender (2),\\ TC (8)\end{tabular} \\ 
\hline
\begin{tabular}[c]{@{}l@{}}HbA1c and\\Diabetes\end{tabular} & \begin{tabular}[c]{@{}l@{}}Age (3), gender (2), \\Diabetes (2) BMI (3), \\HbA1c (6), EQV5 (5)\end{tabular} & \begin{tabular}[c]{@{}l@{}}BMI (3), \\HbA1C (6),\\Diabetes (2)\end{tabular} & \begin{tabular}[c]{@{}l@{}}HbA1C (6),\\Diabetes (2)\end{tabular} \\ 
\hline
\begin{tabular}[c]{@{}l@{}}Smoking \\status\end{tabular} & \begin{tabular}[c]{@{}l@{}}Gender (2), age (4), \\BMI (3), smoking (5), \\EQV5 (5)\end{tabular} & \begin{tabular}[c]{@{}l@{}}Gender (2), age (4),\\smoking (5), \\EQV5 (5)\end{tabular} & \begin{tabular}[c]{@{}l@{}}Gender (2), \\smoking (5), \\EQV5 (5)\end{tabular} \\
\hline
\end{tabular}
\caption{Variables used for matching synthetic individuals to observed individuals in the Markov nonparametric sampling regression method}
\label{tab:match_table}
\end{table}

\end{document}